# Learning Continuous-Time Social Network Dynamics

**Yu Fan**
University of California, Riverside
yfan@cs.ucr.edu

**Christian R. Shelton**
University of California, Riverside
cshelton@cs.ucr.edu

## Abstract

We demonstrate that a number of sociology models for social network dynamics can be viewed as continuous time Bayesian networks (CTBNs). A sampling-based approximate inference method for CTBNs can be used as the basis of an expectation-maximization procedure that achieves better accuracy in estimating the parameters of the model than the standard method of moments algorithm from the sociology literature. We extend the existing social network models to allow for indirect and asynchronous observations of the links. A Markov chain Monte Carlo sampling algorithm for this new model permits estimation and inference. We provide results on both a synthetic network (for verification) and real social network data.

## 1 Introduction

Social networks, which represent the relationships (such as friendship or co-authorship) among actors (such as individuals or companies), have been studied for decades. However, the majority of the existing works model social networks using static or discrete time models. As social networks evolve asynchronously, a continuous-time model can provide more flexibility and higher fidelity in modeling such networks. In addition, the characteristic attributes of the actors and the network's structure (both time-variant) may depend on each other. For example, people who have the same interests are likely to become friends and friends are likely to influence each other's interests. Such effects should be considered when modeling the social networks. Recently, Snijders (2005) provided an actor-oriented model which considers the whole social network evolution as a continuous-time Markov process. Snijders et al. (2007) extended the actor-oriented model to the network-attribute co-evolution model which added effects between the network structure and the actors' attributes.

Due to the large number of variables and dynamic relationships among the variables, exact inference is intractable for learning and reasoning in such systems. Other approximate inference methods such as loopy belief propagation are also hard to implement because of the key role that context sensitive independencies play in the model. Sampling-based algorithms are one applicable method for reasoning in such systems. Parametric estimation in Snijders (2005) and Snijders et al. (2007) was implemented using a forward sampling method. However, this method can only handle data that completely specifies all variables at discrete time instants. The samples generated by their forward sampling algorithm are not entirely consistent with the observations and the information provided by the observations is only partially used during the learning procedure.

Parametric estimation for the network-attribute co-evolution model requires observations of the network as it evolves (at least three snapshots at different times). Usually, direct observation of a social network is very expensive. The scarceness of the data could result in inaccurate estimation of the model. Alternatively, other observations such as communication events among people (emails and instant messages), can reflect people's relationship indirectly. More importantly, they are easier to collect. Utilization of these data can greatly help us to study the dynamics of social networks.

Continuous time Bayesian networks (CTBNs) (Nodelman et al., 2002) model asynchronous stochastic systems' independencies among discrete-valued processes. In this paper, we first show the relation between the CTBN model and the social network dynamics model of Snijders et al. (2007). We then demonstrate a parameter estimation method based on the CTBN importance sampling of Fan and Shelton (2008). This method is a full sampling-based inference method and can handle any type of observation data including asynchronous, partial, and duration observations. We then present a hidden social network dynamics model in which indirect observations such as emails events among people can be utilized. We develop a Markov chain Monte Carlo (MCMC) sampling algorithm for the model and apply it to parameter estimation. Our methods not only im-prove upon previous parameter estimation methods for social network dynamics, they also extend their range to deal with more flexible data sources.

### 1.1 Previous Work

The idea of using a continuous-time model for social networks is not new. The reciprocity model in Wasserman (1979) considers the evolution of the links between any two actors in a network as a continuous-time Markov process with four states. The transition rate of the state is time homogeneous and assumed to be independent of all the other links in the network. The popularity model in Wasserman (1980) assumes that the transition rate of a link depends on the number of incoming links of the actor that the transition link points to. Thus, the transition rate is time-variant and the state of popular actors who have more incoming links may change faster. These two models are computationally efficient but have limited ability to represent some common properties of social networks, such as transitivity.

The actor-oriented model, proposed in Snijders (2005), is an extension of the reciprocity model. It allows the probability that a link changes to depend on the entire network structure. The model can be simulated using forward sampling and the method of moments (Bowman & Shenton, 1985) is used to estimate the parameters. Steglich et al. (2006) shows an application of this model. An alternative Bayesian-based parameter estimation method is implemented in Koskinen and Snijders (2007).

## 2 Background

We briefly review the network-attribute co-evolution model (Snijders et al., 2007), the continuous time Bayesian network model (Nodelman et al., 2002), and the importance sampling algorithm for CTBNs (Fan & Shelton, 2008).

### 2.1 Continuous-Time Social Network Model

The evolution of a social network depends not only on the network structure, but also on the characteristics of the actors. At the same time, the values of the actors' attributes are also influenced by the structure of the network. In the network-attribute co-evolution model, actors change their connections and attributes to maximize a utility function.

**Continuous-Time Markov Process** Let $X$ be a continuous-time, finite-state, homogeneous Markov process with $n$ states $\{x_1, \ldots, x_n\}$. The behavior of $X$ is described by the initial distribution $P_X^0$ and the intensity matrix

$$\mathbf{Q}_X = \begin{bmatrix} -q_{x_1} & q_{x_1 x_2} & \cdots & q_{x_1 x_n} \\ q_{x_2 x_1} & -q_{x_2} & \cdots & q_{x_2 x_n} \\ \vdots & \vdots & \ddots & \vdots \\ q_{x_n x_1} & q_{x_n x_2} & \cdots & -q_{x_n} \end{bmatrix},$$

where $q_{x_i x_j}$ is the intensity with which $X$ transitions from $x_i$ to $x_j$ and $q_{x_i} = \sum_{j \neq i} q_{x_i x_j}$. The intensity matrix $\mathbf{Q}_X$ is time-invariant. Given $\mathbf{Q}_X$, the amount of time $X$ stays at $x_i$ follows an exponential distribution with parameter $q_{x_i}$. That is, the probability density of $X$ remaining at $x_i$ for duration $t$ is $q_{x_i} \exp(-q_{x_i} t)$. The probability that $X$ transitions from $x_i$ to $x_j$ is $\theta_{x_i x_j} = q_{x_i x_j}/q_{x_i}$.

**Network-attribute Co-evolution Model** There are two types of variables in the network-attribute co-evolution model: the evolving pair-wise relationships among the $N$ actors and the $H \geq 1$ discrete-valued attributes[1] for each actor. The number of actors is fixed during the entire process. At any time $t$, the relation links can be described as a directed graph, which is represented by an $N \times N$ adjacency matrix $Y(t)$, where $Y_{ij}(t)$ represents the relation directed from actor $i$ to actor $j$ ($i, j = 1, \ldots, N$). $Y_{ij}(t) = 1$ if there is a tie from actor $i$ to $j$ and $Y_{ij}(t) = 0$ otherwise. Self relations are not considered in the network. The actors' attributes at $t$ can be represented by $H$ integer vectors $Z_h(t)$ of size $N$, where $Z_{hi}(t)$ denotes the value of actor $i$ on attribute $h$. Therefore, the network-attribute co-evolution is modeled using the stochastic process $X(t) = (Y(t), Z_1(t), \ldots, Z_H(t))$.

The network-attribute co-evolution model assumes that the process $X(t)$ is a continuous-time Markov process. The evolution of the network is modeled as actors making decisions to maximize their satisfaction with the network: an actor may choose to add or remove an outgoing tie, or change the value of one attribute in order to (approximately and locally) maximize a utility function. At any time $t$, given the current state $X(t)$, the decisions made by the actors are conditionally independent. It is further assumed that when making a decision, the actor can only change one outgoing tie or change one attribute value. Because of the continuous-time nature, no two events may occur at exactly the same time. Thus, at any time, only one actor can add or remove an outgoing tie or increase or decrease one value unit of an attribute. For each actor $i$, the times between two network changes and between two attribute decisions are exponentially distributed with parameters $\lambda_i^n$ and $\lambda_i^a$, called rate functions. Usually, they are assumed to be constant values.

At a transition point, the actor chooses to add or remove a tie to maximize the value of the network utility function $f_i^n(\beta^n, \boldsymbol{y}\langle i, j\rangle, \boldsymbol{z}) + \epsilon^n$, or to change the value of an attribute to maximize the attribute utility function $f_i^a(\beta^a, \boldsymbol{y}, \boldsymbol{z}_h\langle i, \delta\rangle) + \epsilon^a$, where $f_i^n(\beta^n, \boldsymbol{y}\langle i, j\rangle, \boldsymbol{z})$ and $f_i^a(\beta^a, \boldsymbol{y}, \boldsymbol{z}_h\langle i, \delta\rangle)$ are the objective functions for network and attribute decisions respectively. $\boldsymbol{y}$ and $\boldsymbol{z}$ are the current states of the network and attributes. $\boldsymbol{y}\langle i, j\rangle$ denotes the state of the network after the tie from $i$ to $j$ changes.

---

[1] Snijders et al. (2007) call the attributes of the actors "behaviors." We call them "attributes" to avoid confusion with the dynamic behavior of the model.

$z_h \langle i, \delta \rangle$ denotes the state of the attribute after actor $i$ changes the attribute $z_h$ by $\delta$, where $\delta \in \{-1, +1\}$. Both objective functions are modeled as a weighted sum of effects (features) that depend on the topology of the network and the attribute values. The functions have the form

$$f_i(\beta, \boldsymbol{y}, \boldsymbol{z}) = \sum_{k=1}^{L} \beta_k s_{ik}(\boldsymbol{y}, \boldsymbol{z})$$

where $s_{ik}(\boldsymbol{y}, \boldsymbol{z})$ is an effect that expresses a property of the network structure and the attribute values from the view of actor $i$, and $L$ is the number of effects the actor considers. For example, we can choose $s_{ik}(\boldsymbol{y}, \boldsymbol{z})$ to be the "density effect" defined as the number of out-going ties from $i$, or the "attribute tendency" of actor $i$, which is the current attribute value of $i$.

$\epsilon^n$ and $\epsilon^a$ are random noise. Following Snijders et al. (2007), we set them to be Gumbel distributions with mean 0 and scale parameter 1. Then the transition probability of actor $i$ changing the tie to $j$ and actor $i$ changing the value of $z_h$ by $\delta$ are

$$P(\boldsymbol{y}\langle i,j \rangle | \boldsymbol{y}, \boldsymbol{z}) = \frac{\exp(f_i^n(\beta^n, \boldsymbol{y}\langle i,j \rangle, \boldsymbol{z}))}{\sum_{k \neq i} \exp(f_i^n(\beta^n, \boldsymbol{y}\langle i,k \rangle, \boldsymbol{z}))} \quad (1)$$

$$P(\boldsymbol{z}_h\langle i,\delta \rangle | \boldsymbol{y}, \boldsymbol{z}) = \frac{\exp(f_i^a(\beta^a, \boldsymbol{y}, \boldsymbol{z}_h\langle i,\delta \rangle))}{\sum_{\delta} \exp(f_i^a(\beta^a, \boldsymbol{y}, \boldsymbol{z}_h\langle i,\delta \rangle))} \; . \quad (2)$$

Given the rate functions and the transition probabilities, the intensity matrix of the continuous Markov process $X(t)$ can be written as

$$q_{\boldsymbol{x}, \hat{\boldsymbol{x}}} = \begin{cases} \lambda_i^n P(\boldsymbol{y}\langle i,j \rangle | \boldsymbol{y}, \boldsymbol{z}) & \text{if } \hat{\boldsymbol{x}} = (\boldsymbol{y}\langle i,j \rangle, \boldsymbol{z}) \\ \lambda_i^a P(\boldsymbol{z}_h\langle i,\delta \rangle | \boldsymbol{y}, \boldsymbol{z}) & \text{if } \hat{\boldsymbol{x}} = (\boldsymbol{y}, \boldsymbol{z}_h\langle i,\delta \rangle) \\ -(\lambda_i^n + \lambda_i^a) & \text{if } \hat{\boldsymbol{x}} = \boldsymbol{x} \\ 0 & \text{otherwise.} \end{cases} \quad (3)$$

**Parameter Estimation** Snijders et al. (2007) assume that observations are only available at discrete time points $t_1 < t_2 < \cdots < t_M$, where $M \geq 3$. The parameters $\alpha = (\lambda^n, \lambda^a, \beta^n, \beta^a)$ are estimated based on $M$ network observations $y(t_1), \ldots, y(t_M)$ and attribute observations $z(t_1), \ldots, z(t_M)$. Parameters are estimated using the method of moments (MoM) (Bowman & Shenton, 1985).

MoM estimates the parameters such that the expected values of some statistics $D(\boldsymbol{y}, \boldsymbol{z})$ under the estimated parameter are equal to the observed values. The statistics used by Snijders et al. (2007) for each parameter are as follows.

| Parameter | $D(\boldsymbol{y}, \boldsymbol{z})$ |
|---|---|
| $\lambda^n$ | $\sum_{i,j} |y_{ij}(t_{m-1}) - y_{ij}(t_m)|$ |
| $\lambda^a$ | $\sum_{h,i} |z_{hi}(t_{m-1}) - z_{hi}(t_m)|$ |
| $\beta_k^n$ | $\sum_i s_{ik}^n(\boldsymbol{y}(t_m), \boldsymbol{z}(t_{m-1}))$ |
| $\beta_k^a$ | $\sum_i s_{ik}^a(\boldsymbol{y}(t_{m-1}), \boldsymbol{z}(t_m))$ |

Their estimation algorithm uses the Newton-Raphson method starting with random parameters. In each iteration, the expectation of the statistic values are calculated using forward sampling between two time points $t_m$ and $t_{m+1}$.

## 2.2 Continuous Time Bayesian Networks

Continuous time Bayesian networks (CTBNs) (Nodelman et al., 2002) are factored representations of general Markov processes. We review them here and then relate the network-attribute co-evolution model to them in Section 3.

Continuous time Bayesian networks factorize a homogeneous continuous Markov process into local variables. The intensities for each variable $X$ are described using a conditional intensity matrix (*CIM*) $\mathbf{Q}_{X|\mathbf{U}}$, which is defined as a set of intensity matrices $\mathbf{Q}_{X|\mathbf{u}}$, one for each instantiation $\mathbf{u}$ of the variable set $\mathbf{U}$. The evolution of $X$ depends instantaneously on the values of the variables in $\mathbf{U}$. A *continuous time Bayesian network* $\mathcal{N}$ over $\boldsymbol{X}$ consists of two components: an *initial distribution* $P_{\boldsymbol{X}}^0$, specified as a Bayesian network $\mathcal{B}$ over $\boldsymbol{X}$, and a *continuous transition model*, specified using a directed (possibly cyclic) graph $\mathcal{G}$ whose nodes are $X \in \boldsymbol{X}$. Each variable $X \in \boldsymbol{X}$ is associated with a conditional intensity matrix, $\mathbf{Q}_{X|\mathbf{U}_X}$, where $\mathbf{U}_X$ denotes the parents of $X$ in $\mathcal{G}$,

**Forward sampling semantics for a CTBN** The dynamics of a CTBN can be described using a forward sampling algorithm. For each variable $X \in \boldsymbol{X}$, it maintains $x(t)$ — the state of $X$ at time $t$ — and *Time*$(X)$ — the next potential transition time for $X$, which is sampled from an exponential distribution. The algorithm adds transitions one at a time, advancing $t$ to the next earliest variable transition. The next state is sampled from a multinomial distribution. When a variable $X$ (or one of its parents) undergoes a transition, *Time*$(X)$ is resampled.

**Likelihood and Sufficient Statistics** A CTBN defines a probability density over trajectories $\sigma$ of a set of variables $\boldsymbol{X}$. It is a member of the exponential family, so one way to describe the distribution is to use the sufficient statistics of $\sigma$ (Nodelman et al., 2003). Let $T[x|\mathbf{u}]$ be the amount of time $X = x$ while $\mathbf{U}_X = \mathbf{u}$, and $M[x, x'|\mathbf{u}]$ be the number of transitions from $x$ to $x'$ while $\mathbf{U}_X = \mathbf{u}$. If we let $M[x|\mathbf{u}] = \sum_{x'} M[x, x'|\mathbf{u}]$, the probability density of trajectory $\sigma$ (omitting the starting distribution) is

$$P_{\mathcal{N}}(\sigma) = \prod_{X \in \boldsymbol{X}} L_X(T[X|\mathbf{U}], M[X|\mathbf{U}]) \quad (4)$$

where $L_X(T[X|\mathbf{U}], M[X|\mathbf{U}])$

$$= \prod_{\mathbf{u}} \prod_x \left( q_{x|\mathbf{u}}^{M[x|\mathbf{u}]} \exp(-q_{x|\mathbf{u}} T[x|\mathbf{u}]) \prod_{x' \neq x} \theta_{xx'|\mathbf{u}}^{M[x,x'|\mathbf{u}]} \right)$$

is the local likelihood for variable $X$.

**Importance Sampling** Inference for CTBNs is the task of estimating the distribution over trajectories given some evidence (a trajectory with some missing values or transitions for some variables during some time intervals). Since exact inference is intractable for large networks, approximate inference is often used.

We previously presented an importance sampling algorithm for CTBNs (Fan & Shelton, 2008). Instead of sampling from the target distribution $P_\mathcal{N}$ defined by the CTBN, it samples from a proposal distribution $P'$, which guarantees that all sampled trajectories will conform to the evidence **e**. Each sample $\sigma$ is weighted as $w(\sigma) = \frac{P_\mathcal{N}(\sigma, \mathbf{e})}{P'(\sigma)}$. If we draw a set of samples $\mathcal{D} = \{\sigma[1], \ldots, \sigma[M]\}$ i.i.d. from the proposal distribution, the estimated conditional expectation of a function $f$ given evidence **e** is

$$\hat{\mathbf{E}}_\mathcal{N}[f \mid \mathbf{e}] = \frac{1}{W} \sum_{m=1}^{M} f(\sigma[m]) w(\sigma[m]) \quad (5)$$

where $W = \sum_{m=1}^{M} w(\sigma[m])$.

The proposal distribution is based on forward sampling. At each sample step, the importance sampler adds a new transition to the sampled trajectory and advances time from the current time $t$ to $t + \Delta t$. To ensure variables are consistent with the evidence, it occasionally departs from the regular forward sampling algorithm and "forces" the behavior of one or more variables. The weight contribution for each variable in each step is calculated accordingly.

In each time step from $t$ to $t + \Delta t$, the importance sampling algorithm chooses the proposal distribution for each variable based on the following three cases:

- If there is no upcoming evidence for the variable, or the current state of the variable is the same as the upcoming evidence, the next transition time is sampled from an exponential distribution. The weight contribution for that variable is 1.
- If the behavior of the variable is given according to the evidence, the trajectory of the variable is set to be the same as the evidence in that time interval. The weight contribution is the likelihood of the variable in the interval $[t, t + \Delta t]$. This case corresponds to standard likelihood weighting.
- If the current state of the variable does not agree with the upcoming start of evidence at $t_e$, the next transition time is sampled from a truncated exponential distribution $\frac{q \exp(-q \Delta t)}{1 - \exp(-q(t_e - t))}$, where $q$ is the corresponding intensity for the variable. If the variable is involved in the next transition, the weight contribution is $1 - \exp(-q(t_e - t))$. Otherwise, it is $\frac{1 - \exp(-q(t_e - t))}{1 - \exp(-q(t_e - t - \Delta t))}$.

The weight of the sampled trajectory is the product of the weight contributions of all the variables in each time step.

## 3 Sampling for Learning Social Networks

The method of moments (MoM) parameter estimation method in Snijders et al. (2007) only uses some sufficient statistics of the observation data. Samples generated during the estimation procedure do not fully agree with the observations, which may affect the estimation accuracy. Additionally, calculating the true statistics, $D(\mathbf{y}, \mathbf{z})$, requires that, at each observation point, all the network and attribute values are fully observed. Thus, it is hard for MoM estimation to handle observation data with missing values or evidence about durations, common in real applications.

The network-attribute co-evolution model assumes that the entire state $X(t)$ of the social network is a continuous-time Markov process. We can easily convert the model to a CTBN and apply the importance sampling algorithm to the converted model. The samples generated by importance sampling algorithm are consistent with the observations and we apply maximum likelihood estimation.

### 3.1 Importance Sampling for Network-attribute Co-evolution Model

It is natural to decompose the model into variables $Y_{ij}(t)$ ($i, j = 1, \ldots, N, i \neq j$) and $Z_{hi}(t)$ ($h = 1, \ldots, M, i = 1, \ldots, N$) where $Y_{ij}(t)$ is a binary variable representing the link status from actor $i$ to actor $j$ and $Z_{hi}(t)$ denotes the state of attribute variable of actor $i$. Given the current instantiation $(\mathbf{y}, \mathbf{z})$ of the whole system, the conditional intensity matrix $Q^n_{ij|\mathbf{y}, \mathbf{z}}$ for link variable $y_{ij}(t)$ can be extracted from Equation 3:

$$Q^n_{ij|\mathbf{y}, \mathbf{z}} = \begin{bmatrix} -\lambda^n_i P^0_{i,j} & \lambda^n_i P^0_{i,j} \\ \lambda^n_i P^1_{i,j} & -\lambda^n_i P^1_{i,j} \end{bmatrix}. \quad (6)$$

where $P^k_{i,j} = P(\mathbf{y}\langle i, j \rangle | y_{ij} = k, \mathbf{y}, \mathbf{z})$ for $k = 0, 1$. The conditional intensity matrix $Q^a_{i|\mathbf{y}, \mathbf{z}}$ for attribute variable $Z_{hi}(t)$ can be extracted similarly.

The transition probabilities depend on the utility functions, whose values depend on the current instantiation of the entire system. Independencies among variables only hold for particular assignments to certain other variables (analogous to context sensitive independence for Bayesian networks). For example, if the network utility function's features includes the attribute similarity to connected actors, then the dynamics of link $Y_{ij}$ can potentially depend on all $Z_k$, but at any given instant, only depend on those $Z_k$ for which $Y_{ik} = 1$. If any one of these links $Y_{ik}$ changes, the independencies between $Y_{ij}$ and $\{Z_k\}$ will change. However, the independencies are fixed between any two consecutive transitions. This dynamic CTBN structure prevents efficient use of other approximate CTBN inference algorithms like expectation propagation (Saria et al., 2007). We could apply the Gibbs sampling algorithm of El-Hay et al. (2008), but it must calculate the true posterior density between every two consecutive transitions which can be arbitrarily complex. Sampling from the posterior density involves binary search, which we have found to be too computational costly.

Using the converted CTBN model, we can apply the impor-tance sampling algorithm for the model. This also allows for general evidence patterns beyond complete snapshots.

## 3.2 Maximum Likelihood Estimation

The importance sampling algorithm can generate weighted samples that fully agree with the observations. The log-likelihood of the samples can be calculated using Equation 4 and Equation 5. Therefore, we can use maximum likelihood estimation to learn the parameters $\alpha = (\lambda^n, \lambda^a, \beta^n, \beta^a)$. Since the data are partially observed, we employ the Monte Carlo expectation maximization (MCEM) algorithm (Wei & Tanner, 1990). For this application of EM, the steps are as follows.

**Expectation Step:** Using the current parameters, apply the importance sampling algorithm to generate $m$ weighted samples. Calculate the sufficient statistics and the log-likelihood of the samples.

**Maximization Step:** Using the sufficient statistics and log-likelihood as if they came from the complete data, update parameters. Rate parameters $\lambda_i^n$ and $\lambda_i^a$ are set to be $M^n[i]/T$ and $M^a[i]/T$ respectively, where $M^n[i]$ and $M^a[i]$ are the number of link changes and attribute changes for actor $i$. The weight parameters $\beta^n$ and $\beta^a$ can not be solved analytically from the log-likelihood function. We use conjugate gradient ascent to estimate $\beta$.

Notice that the rate parameters and the weight parameters can be updated separately. To increase the accuracy, we divide the EM iterations into two loops so that the two sets of parameters are estimated alternatingly.

The rate parameters are calculated using the expected number of transitions of the model. Since the time intervals between transitions are sampled from the exponential distribution with the current intensity rate $q$, it is difficult to sample a trajectory with a slower rate. (The algorithm tends to add additional transitions to get the trajectory to agree with the evidence.) Therefore, if the initial rate parameter is larger than the true value, it is unlikely that the EM algorithm will converge to the true rate with a reasonable number of samples. Therefore, instead of sampling transitions using the current intensity $q$, we sample transitions from the exponential and truncated exponential distribution with intensity $q/2$. We adjust the sample weight accordingly.

## 4 Hidden Social Network Dynamics Model

The network-attribute co-evolution model assumes that the adjacency matrix $Y(t)$ can be observed $M \geq 3$ times. Even obtaining one complete observation of the network is very expensive. However, communication events among people, such as emails, instant messages, and phone calls, are easier to collect. We can fully observe these events continuously with lower cost. The occurrence of these events depends on the connection status between people. Thus, although they may not be as accurate, these events indirectly reflect the relationships among people. In such model, the network itself is unobserved (hidden) all the time. We call this model the hidden social network dynamics model.

### 4.1 Model Definition

Let $Y(t)$ be the network of $N$ actors, which evolves in continuous time in the same way as in the network-attribute co-evolution model. (We do not consider the attribute variables. Adding them to the model is straight-forward). Let $O(t)$ be the observations (such as emails or phone calls) among the $N$ actors. $O_{ij}(t) \in O(t)$ $(i, j = 1, \ldots, N, i \neq j)$ is the observation directed from $i$ to actor $j$. We assume that the dynamics of $O_{ij}(t)$ depends only on $Y_{ij}(t)$ and $Y_{ji}(t)$ and it is fully observed as $Y(t)$ evolves. Usually, $O(t)$ is just a set of instantaneous events. There is no "state" for this kind of variable. We use "toggle variables" to model such variables as a continuous-time Markov process. That is, each variable $O_{ij}(t)$ is a binary variable containing two indistinguishable states. The intensities with which the variable transitions in both states are required to be the same. The instantaneous event is represented as the variable flipping between states. We assume that all the event variables share the same set of parameters. That is, the intensity matrix for $O_{ij}(t)$ is

$$Q^{obs}_{ij|Y_{ij}=k,Y_{ji}=l} = \left[ \begin{array}{cc} -q^{obs}_{kl} & q^{obs}_{kl} \\ q^{obs}_{kl} & -q^{obs}_{kl} \end{array} \right]$$

where $k, l \in \{0, 1\}$.

Therefore, the hidden social network dynamic model contains two sets of variables: the network variables $Y(t)$ and the observations $O(t)$. The evolution of the network variables is the same as in the network-attribute co-evolution model. It depends on the network rate $\lambda^n$ and the utility function $f^n$. The dynamics of each observation variable $O_{ij}(t)$ depend only on the state of link variables $Y_{ij}(t)$ and $Y_{ji}(t)$. According to the Markov properties of CTBNs, $O_{ij}(t)$ is independent of all the other variables given the entire trajectory of $Y_{ij}(t)$ and $Y_{ji}(t)$.

### 4.2 Metropolis-Hasting Sampling Algorithm

Due to the large number of variables in the hidden social network dynamic model, exact inference is intractable. Since the model can be naturally converted to a CTBN, we can apply the importance sampling algorithm in Fan and Shelton (2008). However, because only $Y(t)$'s children are observed, the trajectory of $Y(t)$ is sampled blindly with no guidance from the evidence when applying the importance algorithm. Any incorrectly sampled variable can make the entire trajectory be highly unlikely. Given the large sample space for $Y(t)$, the importance sampler generates many trajectories with very small weights.

An alternative method is to use MCMC to guide the sampling toward more likely trajectories. Here we develop a Metropolis-Hasting algorithm (Hastings, 1970) which reuses the importance sampling algorithm. The procedure starts with an arbitrary trajectory that is consistent with the evidence. Then in each iteration, it randomly picks a variable $Y_{ij}(t)$ and replaces the entire trajectory of $Y_{ij}(t)$ using the importance sampling algorithm, fixing the rest of the trajectory as evidence. If the acceptance ratio of the new sampled trajectory is larger than a random number $u$ which is sampled from a $[0,1]$ uniform distribution, the new trajectory is accepted. Otherwise, the old trajectory is kept. Let $\sigma$ be the trajectory from the previous iteration, and $\sigma'$ be the sampled trajectory in the current iteration. Let $P$ be the target distribution given by the model and $P'$ be the proposal distribution used by the importance sampling algorithm. The acceptance ratio of the sampled trajectory $\sigma'$ in the Metropolis-Hasting algorithm is defined to be $r(\sigma,\sigma') = \frac{P(\sigma')P'(\sigma|\sigma')}{P(\sigma)P'(\sigma'|\sigma)}$.

We define $\sigma_Y$ to be the trajectory of $Y(t)$ in $\sigma$, $\sigma_{Y_{ij}}$ to be the trajectory of $Y_{ij}(t)$ in $\sigma$ and $\sigma_{/Y_{ij}}$ to be the trajectory of $Y(t)$ except $Y_{ij}(t)$. We define $\sigma_O$ and $\sigma_{O_{ij}}$ similarly. According to the Markov properties of CTBNs, we have

$$\begin{aligned}
r(\sigma,\sigma') &= \frac{P(\sigma'_Y)P(\sigma_O|\sigma'_Y)}{P(\sigma_Y)P(\sigma_O|\sigma_Y)} \frac{P'(\sigma_Y|\sigma'_Y,\sigma_O)}{P'(\sigma'_Y|\sigma_Y,\sigma_O)} \\
&= \frac{P(\sigma'_Y)}{P(\sigma_Y)} \frac{P(\sigma_{O_{ij}},\sigma_{O_{ji}}|\sigma'_{Y_{ij}},\sigma_{Y_{ji}})}{P(\sigma_{O_{ij}},\sigma_{O_{ji}}|\sigma_{Y_{ij}},\sigma_{Y_{ji}})} \\
&\quad \times \frac{\frac{P(\sigma'_{Y_{ij}}|\sigma_{/Y_{ij}})}{P'(\sigma'_{Y_{ij}}|\sigma_{/Y_{ij}})}}{\frac{P(\sigma_{Y_{ij}}|\sigma_{/Y_{ij}})}{P'(\sigma_{Y_{ij}}|\sigma_{/Y_{ij}})}} \frac{P(\sigma_{Y_{ij}}|\sigma_{/Y_{ij}})}{P(\sigma'_{Y_{ij}}|\sigma_{/Y_{ij}})} \\
&= \frac{L(\sigma'_{/Y_{ij}})}{L(\sigma_{/Y_{ij}})} \frac{L(\sigma_{O_{ij}},\sigma_{O_{ij}}|\sigma'_{Y_{ij}},\sigma_{Y_{ji}})}{L(\sigma_{O_{ij}},\sigma_{O_{ij}}|\sigma_{Y_{ij}},\sigma_{Y_{ji}})} \frac{w(\sigma'_{Y_{ij}})}{w(\sigma_{Y_{ij}})} \quad (7)
\end{aligned}$$

where $L(\cdot)$ is the partial likelihood function and $w(\cdot)$ is the weight contribution of the variable in the importance sampling algorithm.

If the proposal distribution is as described in Section 2.2, $w(\sigma'_{Y_{ij}})$ and $w(\sigma'_{Y_{ij}})$ in Equation 7 are 1 since there is no evidence on $Y_{ij}(t)$. If another proposal distribution is used, the weight contribution should be adjusted accordingly.

### 4.3 Parameter Estimation

To estimate the parameters, we can use the Monte Carlo EM algorithm described in Section 3.2. In the expectation step, we use the Metropolis-Hasting algorithm to generate equally weighted samples. In the maximization step, we update the parameters alternatingly. Calculating $\lambda^n$ and $\beta^n$ is the same as described in Section 3.2. $q^{obs}_{kl}$ in the conditional intensity matrix of $O_{ij}$ can be estimated as

$$q^{obs}_{kl} = \frac{\sum_{i \neq j} M[O_{ij}|Y_{ij}=k, Y_{j,i}=l]}{\sum_{i \neq j} T[O_{ij}|Y_{ij}=k, Y_{ji}=l]}.$$

## 5 Experimental Results

We evaluate the learning algorithm using importance sampling on both synthetic data and real sociological data. We compare the result to the method of moments (MoM) learning algorithm. We also evaluate our learning algorithm using MCMC for the hidden social network dynamics model on a real dataset of emails.

### 5.1 Network-attribute Co-evolution Model

We first built a synthetic social network with 10 people and one time-variant attribute for each person. We assume the rates are homogeneous across people. We consider three effects on the network change rule (features): the density effect (number of outgoing links), the reciprocity effect (number of reciprocity links) and the attribute-related similarity. The attribute utility function considers two effects: tendency and similarity. The utility functions are

$$\begin{aligned}
f^n_i(\beta^n, \boldsymbol{y}, \boldsymbol{z}) &= \sum_j (\beta^n_1 y_{ij} + \beta^n_2 y_{ij} y_{ji} + \beta^n_3 y_{ij} \mathbf{sim}_{ij}) \\
f^a_i(\beta^a, \boldsymbol{y}, \boldsymbol{z}) &= \beta^a_1 z_i + \beta^a_2 \sum_j y_{ij} \mathbf{sim}_{ij} \quad (8)
\end{aligned}$$

where $\mathbf{sim}_{ij} = 1 - |z_i - z_j|/Range(z)$ is the attribute similarity between actors $i$ and $j$. These are the same effects or features as in Snijders et al. (2007). We set $(\beta^n_1, \beta^n_2, \beta^n_3) = (-1, 1.5, 1)$ and $(\beta^a_1, \beta^a_2) = (0.1, 1)$ to generate the synthetic data. Rate parameters for the network and attributes were both set to be 0.5.

We also implemented the MoM learning algorithm in Snijders et al. (2007) and compared it to our method. Note that their algorithm can only deal with evenly-spaced, fully observed point evidence. Therefore, for a comparison, the observation dataset was generated by sampling a full trajectory but only recording the values at regular intervals of $\Delta t$. We randomly sampled another 100 full trajectories as testing data. Learning accuracy was tested by calculating the log-likelihood of the testing trajectories under the estimated models. Figure 1 shows the results for different amounts of data and different values of $\Delta t$. We used 400 samples and averaged across 4 runs (although the values were very stable across runs).

Our algorithm outperforms MoM almost all the time. Its accuracy is much higher than MoM when the number of observation is small. Since samples generated by our algorithm fully agree with the observations, they provide more accurate information about the system, which is valuable when observation data are limited. When the time interval is relatively small and there are enough observation data, our algorithm can accurately estimate the model. As $\Delta t$ increases, the estimation accuracy drops. The expected time between transitions for any given variable is 2 time units. Therefore, when $\Delta t = 8$, it is difficult to estimate the number of changes in the network between observations, thus explaining the poor performance of both algorithms.

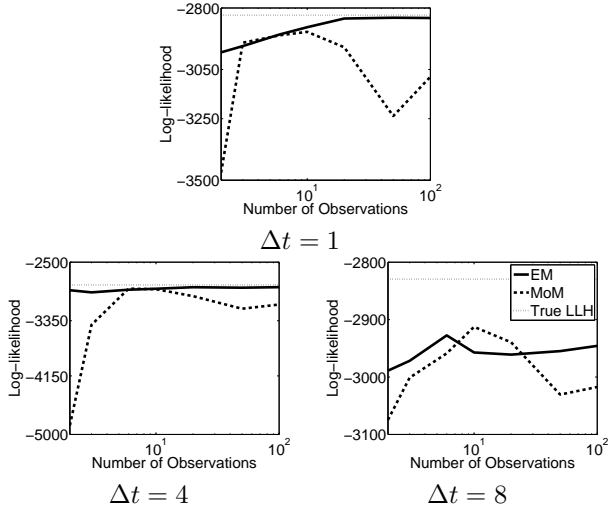

| $\lambda^n$ | Rate/Actor | 0.031 |
| --- | --- | --- |
| $\beta_1^n$ | Density | $-2.362$ |
| $\beta_2^n$ | Reciprocity | 1.210 |
| $\beta_3^n$ | Activity | 0.115 |
| $\beta_4^n$ | Popularity | 0.119 |

| $k, l$ | $q_{kl}^{obs}$ |
| --- | --- |
| 0, 0 | 0.002 |
| 0, 1 | 0.023 |
| 1, 0 | 0.296 |
| 1, 1 | 0.604 |

Figure 3: Estimated parameters for Enron dataset

Figure 1: Log-likelihood of testing data as a function of the number of training data intervals.

| Network Parameter | | | Attribute Parameter | | |
| --- | --- | --- | --- | --- | --- |
| $\lambda^n$ | Rate/Actor | 0.019 | $\lambda^a$ | Rate/Actor | 0.004 |
| $\beta_1^n$ | Density | $-2.39$ | $\beta_1^a$ | Tendency | 0.14 |
| $\beta_2^n$ | Reciprocity | 2.15 | $\beta_2^a$ | Similarity | 1.17 |
| $\beta_3^n$ | Similarity | 0.53 | | | |

Figure 2: Estimated parameters for the 50 girls dataset.

**Real Social Data Example**: We then applied our algorithm to the "50 girls data" from the *Teenage Friends and Lifestyle Study* (Michell & Amos, 1997). The dataset measures the changes in a network of 50 school girls, along with time-variant attributes such as smoking habits and alcohol consumption. The data contains three observations spaced one year apart. In this paper, we concentrate on exploring the effect between the network and the level of alcohol consumption. Alcohol consumption was measured by self-reported questions on a scale ranging from 1 (never) to 5 (more than once a week). We consider the same effects (features) as in the previous experiment. That is, the utility functions have the same format as Equation 8. The true parameters are obviously unknown.

Since this model describes the dynamics of all the links between any two actors in the network and the alcohol attribute for every actor, the model contains 2500 variables; 50 have 5 values and the remainder are binary. No existing exact inference algorithm can handle a system with so many variables. We ran the EM algorithm on this model using 400 samples. The estimated parameters are shown in Figure 2. The time unit was one day. From this result, we can see that, on average, a student changes a friendship every one to two months, and a student's alcohol consumption level remains unchanged for approximately 8 months. Parameters $\beta^a$ indicate that students tend to change their attributes so that they will be similar to their friends. The network parameters $\beta^n$ show that people strongly prefer to build reciprocated connections and they are unlikely to build a connection with someone arbitrary. These parameters seem reasonable and roughly match the rates found by Snijders et al. (2007).

### 5.2 Hidden Social Network Dynamics Model

To evaluate the hidden social network dynamic model, we used the Enron email dataset (Shetty & Adibi, 2004). The dataset contains emails from 151 employees between 1998 and 2002. We preprocessed the data: We only chose emails sent in 2001 since most of the emails were sent in that year. We removed emails whose sender and recipient were the same. Emails that were sent to many recipients were usually general notices such as a reminder of a presentation at certain time and not indicative of a working relationship. If the number of recipients in an email is larger than a threshold $t_r$, we filtered out that email. In our experiment, we set $t_r$ be 5. For the rest of the emails which were sent to multiple recipients, we treated them as multiple single-recipient e-mails and randomly jittered the sent times. We took the intersection of people who sent at least 100 emails in 2001 and people who received at least 100 emails in 2001 as the set of the actors in our model. Emails among these actors were used as our observation. After the process, we obtained a dataset containing 6738 emails for 31 people. We set the time unit to be one day in this experiment.

We considered four effects on the network utility function: the density effect, the reciprocity effect, the activity effect and the popularity effect. Therefore, the utility function is

$$f^n(\beta^n, \boldsymbol{y}) = \sum_j (\beta_1^n y_{ij} + \beta_2^n y_{ij} y_{ji} + \beta_3^n y_{ij} \sum_k y_{jk} + \beta_4^n y_{ij} \sum_k y_{kj}) \ .$$

We ran Monte Carlo EM using the Metropolis-Hasting sampler with 400 samples for each iteration. We took one sample for every 1000 trajectories. Before starting the EM iterations, we sampled 10000 samples as the "burn-in" iterations. In each EM iteration, we used the last trajectory in previous iteration as our initial sample and used 1000 samples as the "burn-in" iterations. We randomly picked the initial parameters for the model and ran the EM algorithm 5 times. All the 5 runs had very similar results. We took the average for each parameter as our learned parameters. The learned parameters are listed in Figure 3.

We can see that, on average, a person would consider changing a working relationship about every month. When choosing the connection to change, the person is very unlikely to build a random connection ($\beta_1^n = -2.1$) and prefers to build reciprocated connections ($\beta_2^n = 1.5$). These results are very similar to what we learned from the "50 girls dataset." The number of popularity and activity of an actor has a positive effect ($\beta_3^n$ and $\beta_4^n$ are multiplied by the number of connections to or from an actor, so they can be smaller than $\beta_2^n$ and still have similar impact). On average, a person $A$ will send an email to another person $B$ about every 3 days if there is a connection from $A$ to $B$. If there is a reciprocated connection, $A$ will send an email to $B$ at least every 2 days. If there is no connection between them, it is unlikely that $A$ will send an email to $B$.

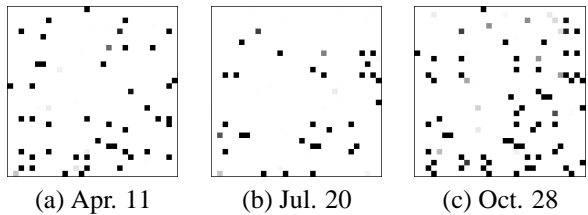

(a) Apr. 11   (b) Jul. 20   (c) Oct. 28

Figure 4: Enron adjacency matrix at different times.

We used the learned parameters as the true parameters of the model. Starting with a random trajectory, we ran our MCMC algorithm for 1,000,000 "burn in" iterations. Then we drew a sample every 1000 iterations. We repeated this until we sampled 1000 trajectories. Using the 1000 trajectories, we calculated the probability $P(Y_{ij}(t) = 1)$, for $i, j = 1, \ldots, 31$, $i \neq j$. Figure 4 shows the network structures as matrices at three different times throughout the year. Darker spots represents higher connection probability. Since the Enron dataset represents working groups changing over one year, we can see that the three matrices are different, showing that the links among people are dynamic processes, but there are some stable connections.

## 6 Conclusion

We provide a sampling-based learning algorithm for continuous-time social network models and provide results for a model with 2500 variables. We also provide a hidden social network dynamics model in which indirect observation of the network can be used and develop an MCMC sampling algorithm for it. Our method is simple and easy to implement. The idea of the algorithm is general and can be easily extended to other continuous-time systems. For social networks, we provide a learning method that is better than the previous method of moments estimation, particularly when data is scarce (a common occurrence in sociology studies). Furthermore, we present a new model that can deal with other type of datasets which previous social network estimation methods cannot. This greatly extends the types of social dynamic data that can be analyzed.

**Acknowledgment** This research was sponsored by AFOSR YIP award FA9550-07-1-0076 and DARPA award HR0011-09-1-0030.

## References

Bowman, K., & Shenton, L. (1985). Method of moments. *Encyclopedia of Statistical Sciences*, 5, 467–473.

El-Hay, T., Friedman, N., & Kupferman, R. (2008). Gibbs sampling in factorized continuous-time Markov processes. *UAI08*.

Fan, Y., & Shelton, C. R. (2008). Sampling for approximate inference in continuous time Bayesian networks. *ISAIM08*.

Hastings, W. K. (1970). Monte Carlo sampling methods using Markov chains and their applications. *Biometrika*, 57, 97–109.

Koskinen, J. H., & Snijders, T. A. (2007). Bayesian inference for dynamic social network data. *Journal of Statistical Planning and Inference*, 137, 3930–3938.

Michell, L., & Amos, A. (1997). Teenage friends and lifestyle study dataset. http://www.stats.ox.ac.uk/~snijders/siena/siena_datasets.htm.

Nodelman, U., Shelton, C. R., & Koller, D. (2002). Continuous time Bayesian networks. *UAI02*.

Nodelman, U., Shelton, C. R., & Koller, D. (2003). Learning continuous time Bayesian networks. *UAI03*.

Saria, S., Nodelman, U., & Koller, D. (2007). Reasoning at the right time granularity. *UAI07*.

Shetty, J., & Adibi, J. (2004). The Enron email dataset database schema and brief statistical report. http://www.isi.edu/~adibi/Enron/Enron.htm.

Snijders, T. A. (2005). *Models for longitudinal network data*, chapter 11. New York: Cambridge Univ. Press.

Snijders, T. A., Steglich, C. E., & Schweinberger, M. (2007). Modeling the co-evolution of networks and behavior. In *Longitudinal models in the behavioral and related sciences*, chapter 4. Lawrence Erlbaum.

Steglich, C., Snijders, T. A. B., & West, P. (2006). Applying SIENA: An illustrative analysis of the co-evolution of adolescents' friendship networks, taste in music, and alcohol consumption. *Methodology*, 2, 48–56.

Wasserman, S. (1979). A stochastic model for directed graphs with transition rates determined by reciprocity. In K. Schuessler (Ed.), *Sociological methodology*, 392–412. Jossy-Bass.

Wasserman, S. (1980). Analyzing social networks as stochastic processes. *J. Am. Stat. Assn.*, 75, 280–294.

Wei, G. C. G., & Tanner, M. A. (1990). A Monte Carlo implementation of the EM algorithm and the poor man's data augmentation algorithms. *J. Am. Stat. Assn.*, 85, 699–704.